\documentclass[a4paper, 11pt]{amsart}
\usepackage{graphicx}
\usepackage{amsfonts}
\usepackage{amsmath}
\usepackage{amssymb}
\usepackage{hyperref}

\begin{document}

\title{The Poincar\'e half-plane for informationally complete POVMs
}

\author{Michel Planat$\dag$}
\address{$\dag$ Universit\'e de Bourgogne/Franche-Comt\'e, Institut FEMTO-ST CNRS UMR 6174, 15 B Avenue des Montboucons, F-25044 Besan\c con, France.}
\email{michel.planat@femto-st.fr}



\begin{abstract}

It has been shown that classes of (minimal asymmetric) informationally complete POVMs in dimension $d$ can be built using the multiparticle Pauli group acting on appropriate fiducial states [M. Planat and Z. Gedik, R. Soc. open sci. 4, 170387 (2017)]. The latter states may also be derived starting from the Poincar\'e upper half-plane model $\mathbb{H}$. For doing this, one translates the congruence (or non-congruence) subgroups of index $d$ of the modular group into groups of permutation gates whose some of the eigenstates are the seeked fiducials. The structure of some IC-POVMs is found to be intimately related to the Kochen-Specker theorem.

\end{abstract}

\maketitle

\vspace*{-.5cm}

\normalsize
\vspace*{.5cm}

\begin{center}
{\it Nulle parole ne trouve \\
une branche o\`u se poser}\footnote{
{\it No words find a branch where to land} \cite{Guesquier}.}
\end{center}

\section{Introduction}

{\it Out of nothing I have created a strange new universe} wrote Janos Bolyai to his father in 1823. Half a century later, Felix Klein published the Erlangen program making explicit the relationship of all geometries to projective geometry via their own groups of symmetries. To illustrate the rise of non Euclidean ideas in this epoch, Let us quote a few words of M. Pasch to  F. Klein in 1891 \cite{DirkSchlimm2013} {\it I am also late in thanking you for sending your essay on non-Euclidean geometry from last year... I also agree with most of what you say in the last two pages of your essay. The content of the axioms comes from observations (intuition as an internal activity is based on remembering what has been observed); the concepts used in the axioms, however, are inexact, and thus so are the axioms themselves. These latter can, however, only be used purely logically if they are presented as being exact. By further working on the axioms and seeking
geometric propositions, we commonly make use of figures, either by drawing them or by ‘imagining’ them... The consideration must be possible
even without the figures, in other words: that which is derived from the figures must already be contained in the axioms, for otherwise the axioms are not complete.} Pasch's axiom (of plane geometry) was used by Hilbert to complete Euclid's axioms. It is related to Pasch's configuration of points and lines (in projective geometry). Incidence geometry, born at the time of Pappus of Alexandria, developed with Desargues (1591-1661), Jacob Steiner (1796-1863), Thomas Kirkman (1806-1895), Gino Fano (1871-1952), David Hilbert (1862-1943) and more recently Jacques Tits (1930-) and Francis Buekenhout (1937-). 

What kind of relation does quantum mechanics maintain with projective geometry? The superposition of states in Hilbert space $\mathcal{H}$ is linear but, due to the probability interpretation of the wave function, a state is not a single vector but a ray, i.e. a one-dimensional subspace of $\mathcal{H}$. As a result, the space of rays is not a linear but a projective space. It is known from Wigner's theorem (1931) that the realization of symmetries for pure states of a QM system is (up to a scalar) a unitary or anti-unitary transformation \cite{Keller2007,Freed2012}, see also \cite{Slomczynski2016,Dang2015} for the relation to POVMs. It is worthwhile to point out that many of the basic configurations of incidence geometry are associated to the commutation relations between Hermitian operators in the (generalized) Pauli group \cite{Planat2017}. Some projective configurations also occur in the structure of informationally complete positive operator valued measures (IC-POVMs) \cite{PlanatRukhsan, PlanatGedik}.

Sec. 2 is a prolegomenon to the modular group $\Gamma$, its finite index subgroups $\Gamma_s$ and the connection to permutation gates considered in previous papers \cite{PlanatRukhsan, PlanatGedik}. Then one deals with the concept of IC-POVMs, the relation to the Pauli group and the occurrence of Kochen-Specker theorem.  The main goal of the paper, worked out in Sec. 3, is the derivation of \lq modular' ICs that follow from the structure of $\Gamma$. In small dimensions $d<10$, triple products of projectors encapsulate recognizable finite geometries (already described in \cite{PlanatGedik}) some of them related to the Kochen-Specker theorem. For $10 \le d \le 27$, many ICs may be constructed thanks to appropriate subgroups $\Gamma_s$. A summary of the results is in Table 1.

\section{Prolegomenon about $\Gamma$ and its relation to IC-POVMs}

\subsection{The modular group $\Gamma$}
{\it The facts of science and, \`a fortiori, its laws are the artificial work of the scientist; science therefore can teach us nothing of the truth; it can only serve us as rule of action} \cite{Poincare}.

A standard non Euclidean geometry consists of the Poincar\'e hyperbolic plane $\mathbb{H}=\{x,y\in \mathbb{R}|y>0\}$ whose symmetry group is the projective (special) linear group $L=PSL(2,\mathbb{R})$ of real M\"obius transformations of $\mathbb{H}$. A discrete subgroup of $L$ is called a Fuchsian group with the modular group $\Gamma=PSL(2,\mathbb{Z})$  as the most celebrated example. Important mathematical objects are the moduli space of elliptic curves, which is the quotient space $\mathbb{H}/\Gamma$, and modular forms that map pair of points of $\mathbb{H}$ up to a weight factor and are also related to elliptic curves (via the 1995 modularity theorem) \cite{Diamond2005}.

The modular group $\Gamma$ acts discontinuously on the extended upper half-plane $\mathbb{H}^*=\mathbb{H}\cup \mathbb{Q} \cup \infty$, i.e. for each $z \in \mathbb{H}^*$ there exists a neighborhood of $z$ with no other element in the orbit of $z$. Thus $\Gamma$ tesselates $\mathbb{H}^*$ with infinitely many copies of the fundamental domain $\mathcal{F}=\{z \in \mathbb{H}~\mbox{with}~ |z|>1, \Re(z)<\frac{1}{2}$. The modular group $\Gamma$ is generated by two transformations $S_\Gamma:z \rightarrow -\frac{1}{z}$ and $T_\Gamma:z \rightarrow z+1$. It can also be represented as the two-generator free group $G=\left\langle e,v |e^2=v^3=1\right\rangle$ using the variable change $e=S_\Gamma$ and $v=S_\Gamma T_\Gamma$.


Some finite index subgroups of $\Gamma$, called congruence subgroups, are obtained by fixing congruence relations on the entries of elements of $\Gamma$. The principal congruence subgroup of level $N$ of $\Gamma$ is the normal subgroup  $\Gamma(N)=\bigl( \begin{smallmatrix} 
  a & b\\
  c & d 
\end{smallmatrix}\bigr) | a,d= \pm1 \mod N ~\mbox{and}~b,c=0 \mod N\}$ whose index is $n^3\Pi_{p|N}(1-\frac{1}{p^2})$, $p$ a prime number. Another important subgroup of $\Gamma$ is the congruence subgroup $\Gamma_0(N)$ of level $N$ defined as the subgroup of upper triangular matrices with entries defined modulo $N$. The index of $\Gamma_0(N)$ is the Dedekind psi function $\psi(N)$. More generally, a congruence subgroup $\Gamma_c$ contains $\Gamma(N)$ for some $N$ and the level of $\Gamma_c$ is the smallest positive $N$ such that $\Gamma(N) \supseteq \Gamma_c$.

Important geometric invariants of a finite index subgroup $\Gamma_s$ of $\Gamma$ (either congruence or not) are the genus, the structure of elliptic points and that of cusps (parabolic points) \cite{Kurth2007}. A fundamental domain $\mathbb{F}_s$ of $\Gamma_s$ in the upper half-plane $\mathbb{H}$ is such that for any $z \in \mathbb{H}$ there is a unique $\gamma \in \Gamma_s$ such that $\gamma(z) \in \mathbb{F}_s$. The subgroup $\Gamma_s$ acts discontinuously on the extended upper half-plane $\mathbb{H}^*$ and tesselates it with infinitely many copies of $\mathbb{F}_s$. Fixed points in $\mathbb{H}$ are elliptic points. An elliptic point of $\Gamma_s$ is a transformation $\gamma \in \Gamma_s$ such that $\gamma(z)=z$ and $\gamma \ne \pm I$, where $I$ is the identity matrix. Elliptic points of $\Gamma_s$ satisfy $|\mbox{tr}(\gamma)|<2$ and their order can only be two or three. Their number are denoted $\nu_2$ and $\nu_3$, respectively. A cusp of $\Gamma_s$ is a fixed point of the extended upper-half plane $\mathbb{H}^*$ such that $|\mbox{tr}(\gamma)|=2$. It can be shown that the action of $\Gamma_s$ partitions $\mathbb{Q}\cup \infty$ into equivalence classes where $q_1 \sim q_2$ if $q_1=\gamma q_2$ for some $\gamma \in \Gamma_s$. These equivalence classes correspond to the cusps of $\Gamma_s$ and the widths of the cusps are the ratios between $\mbox{Stab}_{\Gamma (q)}$ and $\mbox{Stab}_{\Gamma_s (q)}$. The level of $\{q\}$ is the LCM of the cusp widths of $\Gamma$. The structure of cusps is denoted $[\cdots c_i^{w_i}\cdots]$ with $c_i$ the number of cusps of width $W_i$. In Table 1, the signature of a non-congruence subgroup $\Gamma_s$ is represented as NC$(g,N,\nu_2,\nu_3,[\cdots c_i^{w_i}\cdots])$. If the subgroup of $\Gamma$ is a congruence subgroup $\Gamma_c$ the geometric invariants are available in \cite{CumminsPauli}. To conclude this subsection, a Farey symbol for $\Gamma_s$ is a certain finite sequence of rational numbers (they are fractions representing vertices of a fundamental domain of $\Gamma_s$) together with pairing information for the edges between the vertices \cite{Kurth2007}.

Using methods developped in \cite{Kurth2007,CumminsPauli} and implemented in the Sage software \cite{Sage2014}, one can represent a subgroup $G$ of $\Gamma$ either through the permutation representation $P$ of the cosets of $G$ in $\Gamma$ (by making use of the Coxeter-Todd algorithm) or through the modular representation of $P$. Doing this, one arrives at the unexplored relationship of permutation gates of quantum computing, informationally complete POVMs \cite{PlanatRukhsan, PlanatGedik} and the aforementioned modular objects.

\subsection{Minimal informationally complete POVMs and the Pauli group}
{\it In QBism, all the personalist Bayesian properties of probability theory carry over to quantum states; that is, quantum states, like probabilities, are valuations of belief for future experiences} \cite{deBrota2017}.

The paper is a continuation of \cite{PlanatGedik} while restricting to permutation gates stemming from subgroups of $\Gamma$.
Our interest is still the search of minimal informationally complete (IC) POVMs derived from appropriate fiducial states under the action of the (generalized) Pauli group.

A POVM is a collection of positive semi-definite operators $\{E_1,\ldots,E_m\}$ that sum to the identity. In the measurement of a state $\rho$, the $i$-th outcome is obtained with a probability given by the Born rule $p(i)=\mbox{tr}(\rho E_i)$. For a minimal IC-POVM, one needs $d^2$ one-dimensional projectors $\Pi_i=\left|\psi_i\right\rangle \left\langle \psi_i \right|$, with $\Pi_i=d E_i$, such that the rank of the Gram matrix with elements $\mbox{tr}(\Pi_i\Pi_j)$, is precisely $d^2$.

A SIC-POVM obeys the remarkable relation \cite{Renes2004}

\begin{equation}
\left |\left\langle \psi_i|\psi_j \right \rangle \right |^2=\mbox{tr}(\Pi_i\Pi_j)=\frac{d\delta_{ij}+1}{d+1},
\end{equation}
that allows the explicit recovery of the density matrix as in \cite{Fuchs2004}.



In this paper, we discover minimal IC-POVMs (i.e. whose rank of the Gram matrix is $d^2$) and with Hermitian angles $\left |\left\langle \psi_i|\psi_j \right \rangle \right |_{i \ne j} \in A=\{a_1,\ldots,a_l\}$, a discrete set of values of small cardinality $l$. A SIC is equiangular with $|A|=1$ and $a_1=\frac{1}{\sqrt{d+1}}$. The states encountered below are considered to live in a cyclotomic field $\mathbb{F}=\mathbb{Q}[\exp(\frac{2i\pi}{n})]$,
 with $n=\mbox{GCD}(d,r)$, the greatest common divisor of $d$ and $r$, for some $r$. 
The Hermitian angle is defined as $\left |\left\langle \psi_i|\psi_j \right \rangle \right |_{i \ne j}=\left\|(\psi_i,\psi_j)\right\|^{\frac{1}{\mbox{\footnotesize deg}}}$, where $\left\|.\right\|$ means the field norm \cite[p. 162]{Cohen1996} of the pair $(\psi_i,\psi_j)$ in $\mathbb{F}$ and $\mbox{deg}$ is the degree of the extension $\mathbb{F}$ over the rational field $\mathbb{Q}$ \cite{PlanatGedik}.

We construct the relevant IC-POVMs using the covariance with respect to the generalized Pauli group.
Let $d$ be a prime number, the qudit Pauli group is generated by the shift and clock operators as follows
\begin{eqnarray}
&X\left|j \right\rangle = \left|j+1 \mod d \right\rangle \nonumber \\
&Z \left |j \right\rangle=\omega^j \left|j \right\rangle 
\label{eqn1}
\end{eqnarray}
with $\omega=\exp(2i\pi/d)$ a $d$-th root of unity. In dimension $d=2$, $X$ and $Z$ are the Pauli spin matrices $\sigma_x$ and $\sigma_z$. For $N$ particules, one takes the Kronecker product of qudit elements $q$ times.

Stabilizer states are defined as eigenstates of the Pauli group.

\subsubsection*{The single qubit SIC-POVM}

The covariance property under the Pauli group can be illustrated in the two-dimensional case. One can start with the qubit fiducial/magic state
$\left| T\right\rangle=\cos(\beta)\left| 0\right\rangle + \exp{(\frac{i\pi}{4})}\sin(\beta)\left| 1\right\rangle$, $\cos(2\beta)=\frac{1}{\sqrt{3}},$
employed for universal quantum computation \cite{Bravyi2004}. It is defined as the $\omega_3=\exp(\frac{2i\pi}{3})$-eigenstate of the $SH$ matrix [the product of the Hadamard matrix $H$ and the phase gate $S=\bigl( \begin{smallmatrix} 
  1 & 0\\
  0 & i 
\end{smallmatrix}\bigr)$]. Taking the action on $\left| T\right\rangle$  of the four Pauli gates $I$, $X$, $Z$ and $Y$, the corresponding (pure) projectors  $\Pi_i=\left|\psi_i\right\rangle \left\langle \psi_i \right|,i=1\ldots 4$, sum to twice the identity matrix thus building a POVM and the pairwise distinct products satisfy $\left|\left \langle \psi_i \right|\psi_j \right\rangle|^2=\frac{1}{3}$. The four elements $\Pi_i$ form the well known $2$-dimensional SIC-POVM \cite[Sec. 2]{Renes2004}. In contrast, one does not find a POVM attached to the magic state $\left| H\right\rangle=\cos(\frac{\pi}{8})\left| 0\right\rangle + \sin(\frac{\pi}{8})\left| 1\right\rangle$.

\subsection{The Kochen-Specker theorem}
Humans are {\it rats who themselves construct the labyrinth they propose to escape\footnote{Rats qui construisent eux-m\^emes le labyrinthe dont ils se proposent de sortir.}} \cite{Oulipo}.

In a nutshell, the measured value of a quantum observable sometimes depends on which other mutually compatible measurements might be performed. This leads to the concept of quantum contextuality. State independent contextuality is often formulated in terms of the Kochen-Specker (KS) theorem because this theorem is able to guarantee the non-existence of non-contextual hidden variable theories, at least for dimension $d\ge 3$. A non-coloring KS proof consists of a finite set of projectors that cannot be assigned truth values ($1$ for true, $0$ for false) in such a way that (i), one member of each complete orthonormal basis is true and ii), no two-orthogonal (that is mutually compatible) projectors are both true, see \cite{Planat2012,Pavicic2017} and references therein. Remarkably, it will be shown that subsets of projectors within the IC-POVMs below may sometimes be used to derive proofs of KS theorem (in dimensions $4$, $8$ and $9$). Contextuality and negativity of the Wigner function happen to play a fundamental role in universal schemes of quantum computation (e.g. \cite{PlanatRukhsan,deBrota2017,Delfosse2017,deSilva2017}).

\section{Permutation gates from $\Gamma$, fiducial states and informationally complete measurements}

In all the paper, we restrict to IC-POVMs that are built from subgroups of the modular group $\Gamma$ (in contrast to \cite{PlanatGedik} where more general subgroups of the two-generator free groups were also considered).

Using the function \lq G.aspermutation.group()' in Sage \cite{Sage2014}, the permutation representation of the group $G$ is converted into that of the relevant subgroup $\Gamma_s$ of $\Gamma$. If $\Gamma_s$ is congruence, it can be identified in Cummins \& Pauli table \cite{CumminsPauli} from its features, e.g. the genus $g$, the level $N$, the number of elliptic points of order two $\nu_2$, the number of elliptic points of order three $\nu_3$, the number of cusps $c$ of widths $W$ as well as the fractions of the Farey symbol.

\subsection{The three-dimensional Hesse SIC}
The only permutation group that can be used to build a three-dimensional IC-POVM (here a SIC) is the symmetric group $S_3=\left\langle e,v\right\rangle$ with generators $e=(2,3)\equiv \left( \begin{smallmatrix} 
  1 & 0 & 0\\
  0 & 0 & 1\\
	0 & 1 & 0
\end{smallmatrix}\right)$ and $v=(1,2,3)\equiv\left( \begin{smallmatrix} 
  0 & 1 & 0\\
  0 & 0 & 1\\
	1 & 0 & 0
\end{smallmatrix}\right)$, made explicit in terms of the (index $3$) permutation representation and the corresponding permutation gate. Using Sage and the table of congruence subgroups \cite{CumminsPauli} it is straightforward to recognize that $\Gamma_s=\Gamma_0(2)$ whose fundamental domain is pictured in Fig. 1b. In this particular case $\nu_2=1$ [the elliptic point at $z=\frac{1}{2}(1+i)$ is denoted by the symbol *], $\nu_3=0$, the cusps are at $0$ and $\infty$, the fractions are at $0$ and $1$. The subgroup $\Gamma_s=\Gamma_0(2)$ is generated by two transformations $S_{\Gamma_s}:z \rightarrow \frac{z-1}{2z-1}$ and $T_{\Gamma_s}:z \rightarrow z+1$.

The eigenstates of the permutation matrices in $S_3$ can serve as fiducial states for an IC-POVM are qutrits in the classes $(0,1,\pm 1)\equiv \frac{1}{\sqrt{2}}(\left|1\right\rangle \pm \left|2\right\rangle$. Taking the action of the the nine qutrit Pauli matrices, one arrives at the well known Hesse SIC. The Hesse SIC is illustrated in Fig. 1c: the lines of the configuration correspond to projectors whose traces of triple products equal $\pm \frac{1}{8}$ \cite{Tabia2013},\cite[Fig. 1a]{PlanatGedik}. Instead of labeling coordinates as projectors one labels them with the qutrit operators acting on the fiducial state.. 

\begin{figure}[ht]
\includegraphics[width=6cm]{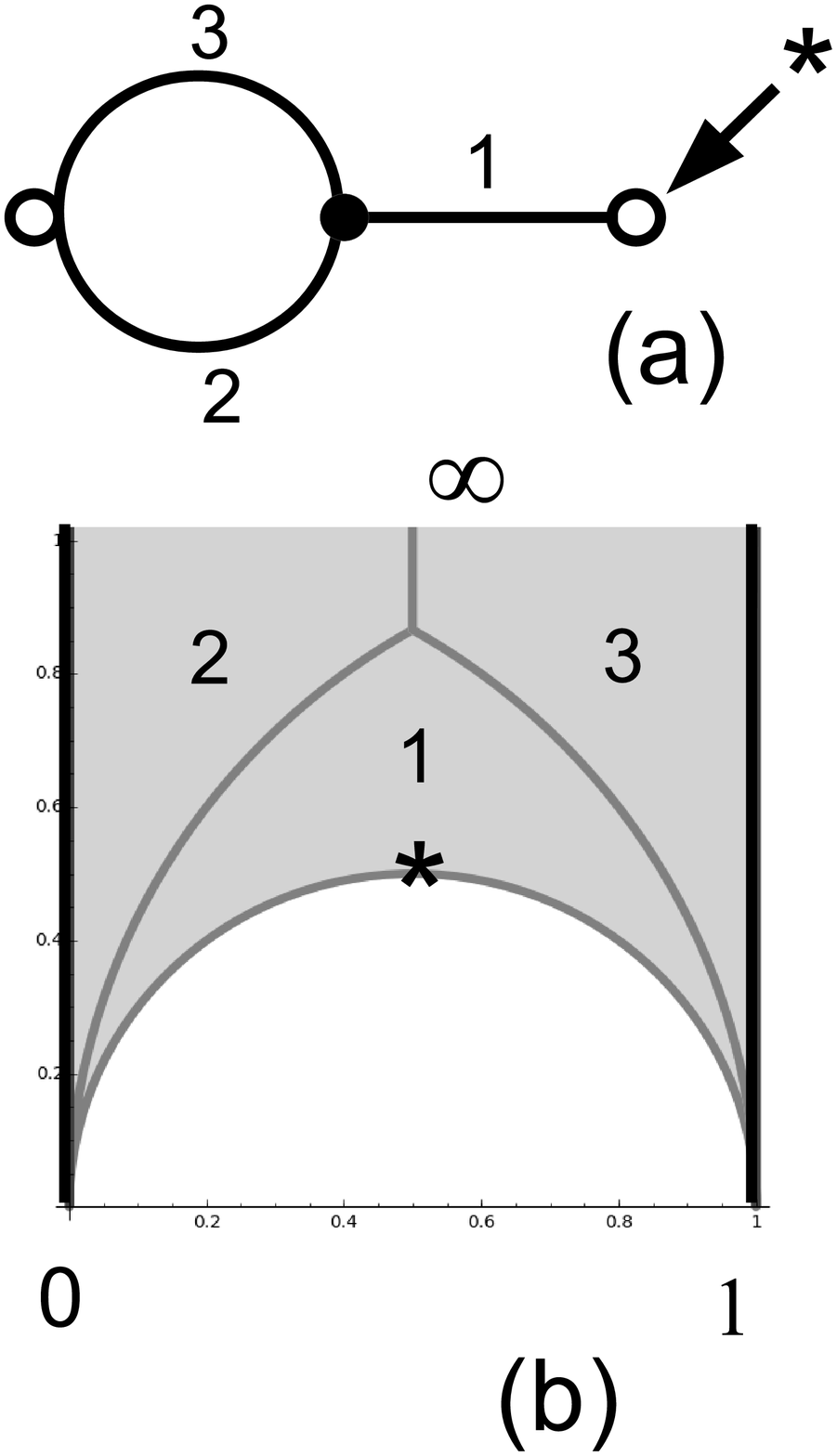}
\includegraphics[width=6cm]{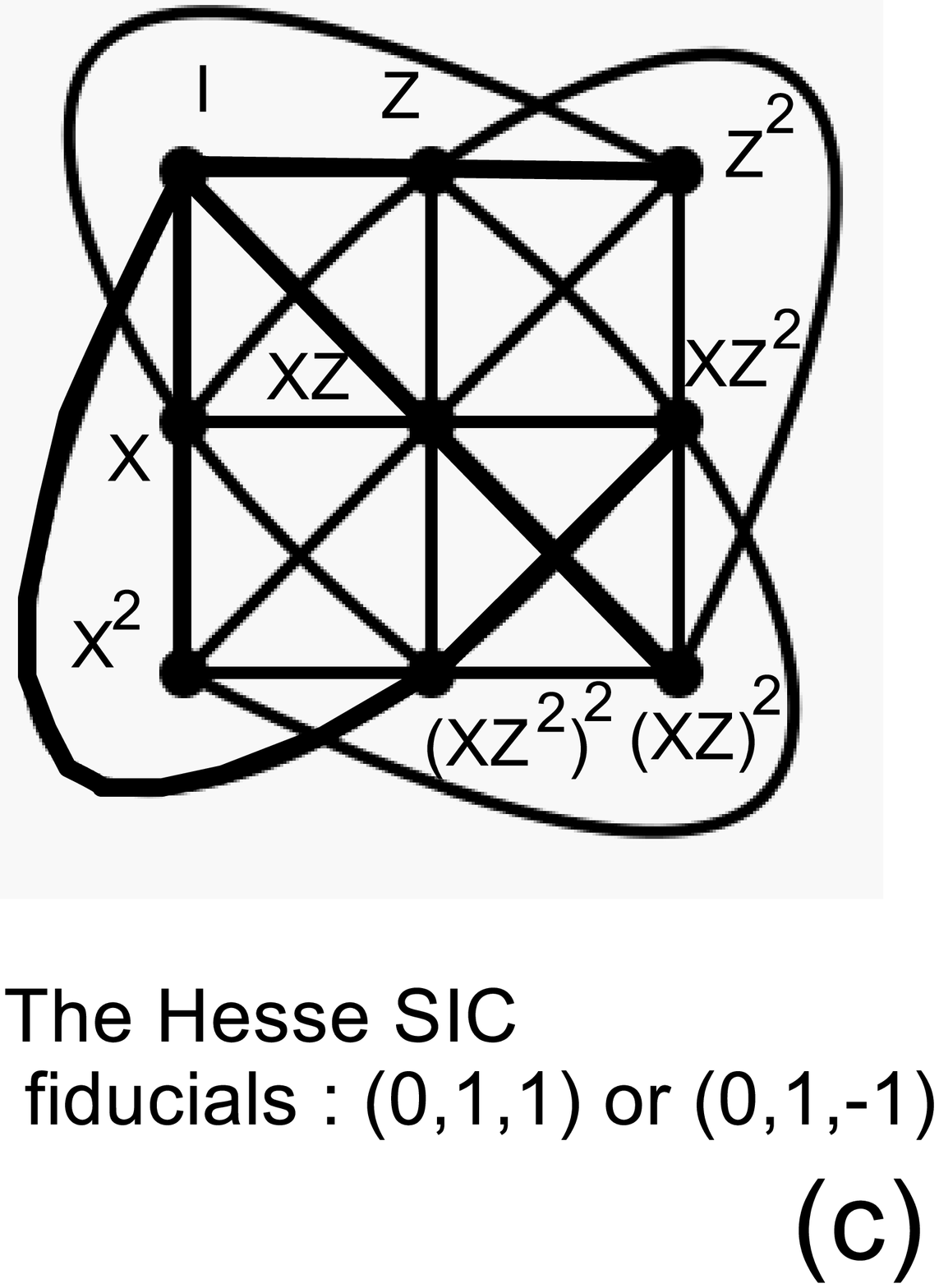}
\caption{Representation of $S_3 \cong \Gamma_0(2)$ as a dessin d'enfant (a) and as the tiling of the fundamental domain (the two thick vertical lines have to identified) (b). The character * denotes the unique elliptic point (of order $2$). The resulting Hesse SIC is in (c).
 }
\end{figure} 

\subsection{The two-qubit IC-POVM}

\begin{figure}[ht]
\includegraphics[width=5cm]{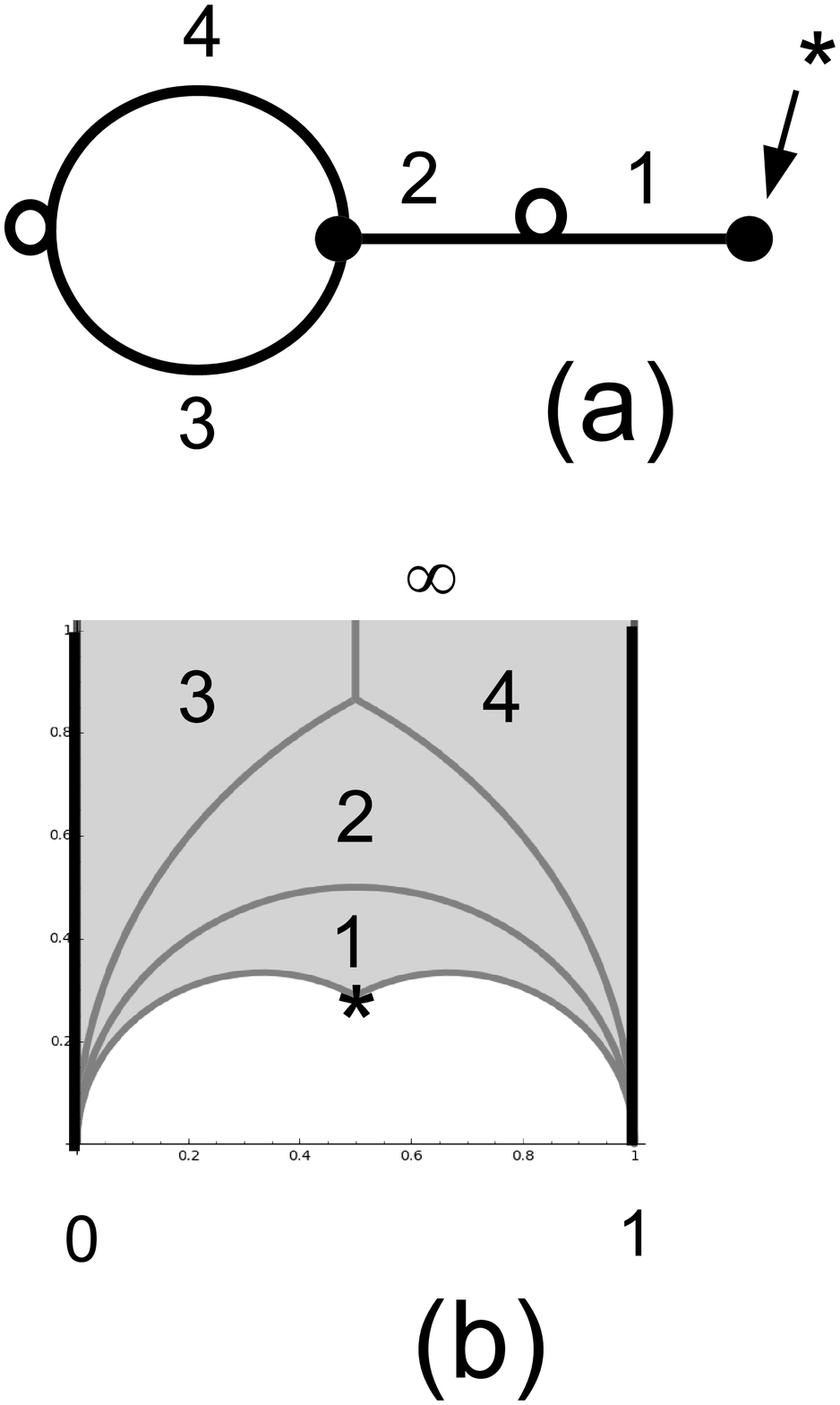}
\includegraphics[width=5cm]{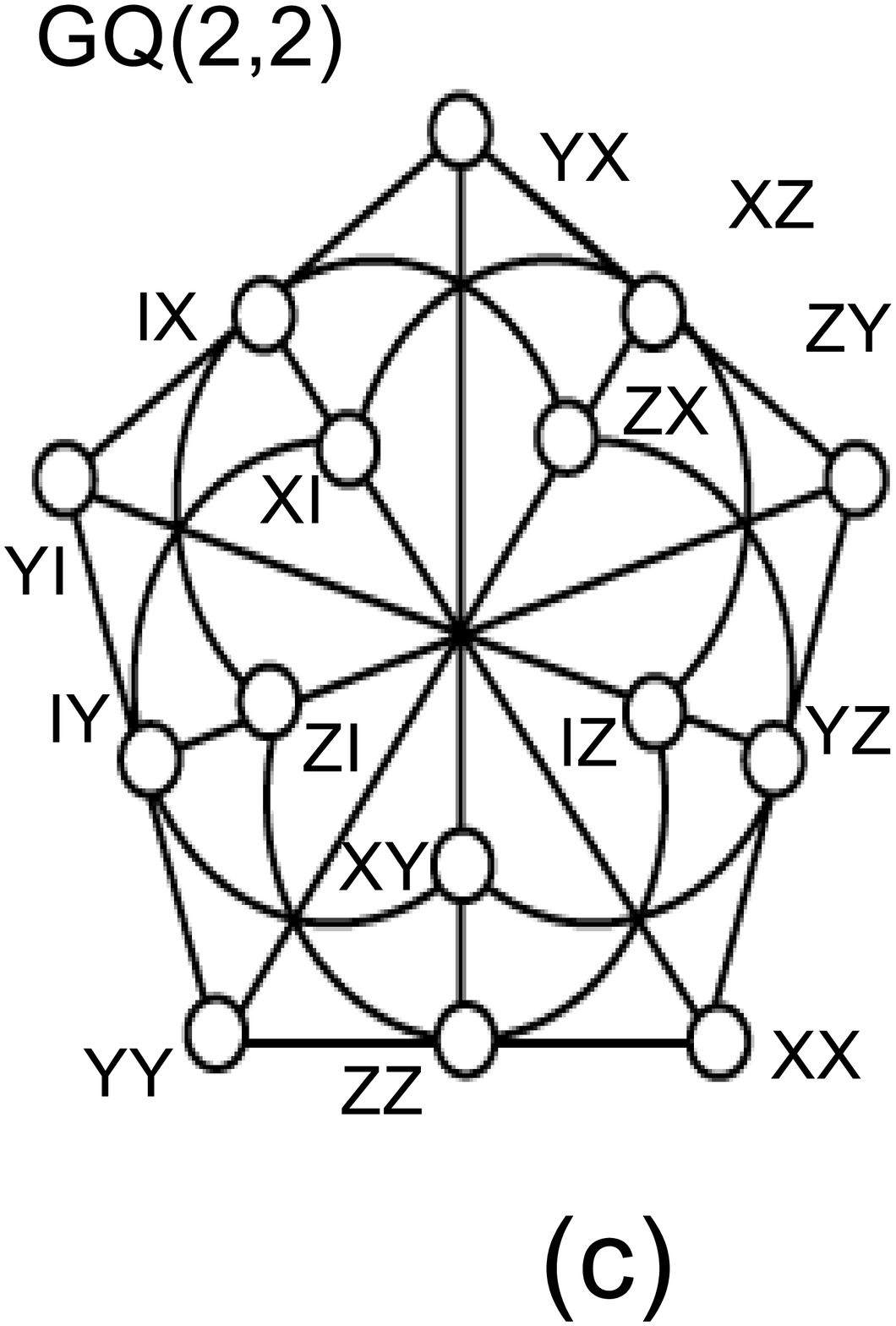}
\includegraphics[width=6cm]{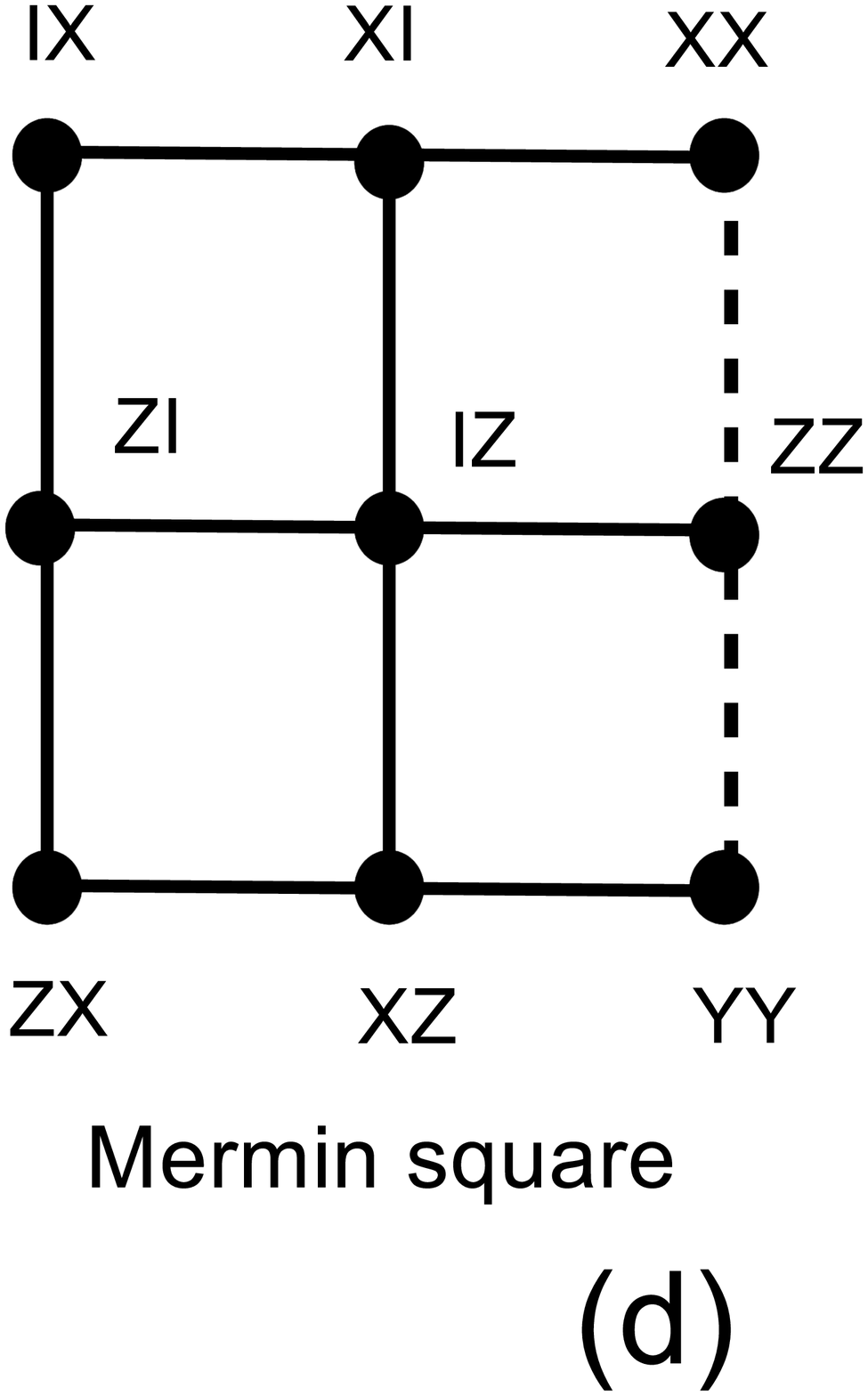}
\caption{Representation of $A_4 \cong \Gamma_0(3)$ as a dessin d'enfant (a) and as the tiling of the fundamental domain (the two thick vertical lines have to identified) (b). The character * denotes the unique elliptic point (of order $3$). The organization of triple products of projectors leads to the generalized quadrangle $GQ(2,2)$ pictured in (c) whose subset is Mermin square (d). Traces of triple products for rows (resp. columns) of Mermin square equal $-\frac{1}{27}$ (resp. $\frac{1}{27}$).
 }
\end{figure}

The smaller permutation group that can be used to build a four-dimensional IC-POVM is the alternating group $A_4=\left\langle e,v\right\rangle$ with generators $e=(1,2)(3,4)\equiv \left(\begin{smallmatrix} 
  0 & 1 & 0 & 0\\
  1 & 0 & 0 & 0\\
	0 & 0 & 0 & 1\\
	0 & 0 & 1 & 0
\end{smallmatrix}\right)$ and $v=(2,3,4)\equiv\left( \begin{smallmatrix} 
  1 & 0 & 0 & 0\\
  0 & 0 & 1 & 0\\
	0 & 0 & 0 & 1\\
	0 & 1 & 0 & 0
\end{smallmatrix}\right)$, made explicit in terms of the (index $4$) permutation representation and the corresponding permutation gate. Using Sage and the table of congruence subgroups \cite{CumminsPauli} one recognizes that $\Gamma_s=\Gamma_0(3)$ whose fundamental domain is pictured in Fig. 2b. In this particular case $\nu_2=0$, $\nu_3=1$ [the elliptic point at $\frac{1}{2}(1+\frac{i}{\sqrt{3}})$ is denoted by the symbol *], the cusps are at $0$ and $\infty$, the fractions are at $0$ and $1$. The subgroup $\Gamma_s=\Gamma_0(3)$ is generated by two transformations $S_{\Gamma_s}:z \rightarrow \frac{z-1}{3z-2}$ and $T_{\Gamma_s}:z \rightarrow z+1$.

The joined eigenstates of the commuting permutation matrices in $S_3$ that can serve as fiducial states for an IC-POVM are of the form
$(0,1,-\omega_6,\omega_6-1)\equiv \frac{1}{\sqrt{3}}(\left|01 \right\rangle-\omega_6\left|10 \right\rangle+(\omega_6 -1)\left|11 \right\rangle)$, with $\omega_6=\exp(\frac{2i\pi}{6})$.
Taking the action of the two-qubit Pauli group on the latter type of state, the corresponding pure projectors sum to four times the identity (to form a POVM) and are independent, with the pairwise distinct products satisfying the dichotomic relation  $\mbox{tr}(\Pi_i \Pi_j)_{i \ne j}=\left|\left \langle \psi_i \right|\psi_j \right\rangle|_{i \ne j}^2
 \in \{\frac{1}{3},\frac{1}{3^2}\}$. Thus the $16$ projectors $\Pi_i$ build an asymmetric informationally complete POVM (see also \cite[Sec. 2]{PlanatGedik}) .

The organization of triple products of projectors whose trace is constant [that is equal to $\frac{1}{9}$ or $\pm \frac{1}{27}$] and simultaneously equal plus or minus the identity matrix $\mathcal{I}$ is shown in Fig. 2c. Instead of labeling coordinates as projectors one labels them with the two-qubit operators acting on the fiducial state. The displayed picture is that of the generalized quadrangle of order two $GQ(2,2)$. It also corresponds to the set of triples of mutually commuting two-qubit operators \cite{Planat2011}. By restricting to triples of projectors whose trace is $\pm \frac{1}{27}$ one identifies the standard Mermin square in Fig. 2d that is known to allow an operator proof of the Kochen-Specker theorem \cite{Planat2012}.

Finally, let us observe that the group $S_4=\left\langle(1,2),(2,3,4)\right \rangle$ may also be used to built the two-qubit IC-POVM as above. It corresponds to the congruence subgroup $4A^0$ in the Cummins-Pauli table: it is of level $4$,  with $\nu_2=2$, $\nu_3=1$ and a single cusp at $\infty$.

\subsection{The five-dimensional equiangular IC-POVM}

There is just one subgroup of index $5$ (up to conjugation) of the two-generator free group isomorphic to $\Gamma$. The organization of cosets defines the alternating group $A_5=\left\langle e,v\right\rangle$ with generators $e=(1,2)(4,5)$ and $v=(2,3,4)$. Using Sage and the table of congruence subgroups \cite{CumminsPauli} one recognizes that $\Gamma_s$ is the congruence subgroup $5A^0$ of level $5$ whose fundamental domain is pictured in Fig. 3b.  
\begin{figure}[ht]
\includegraphics[width=6cm]{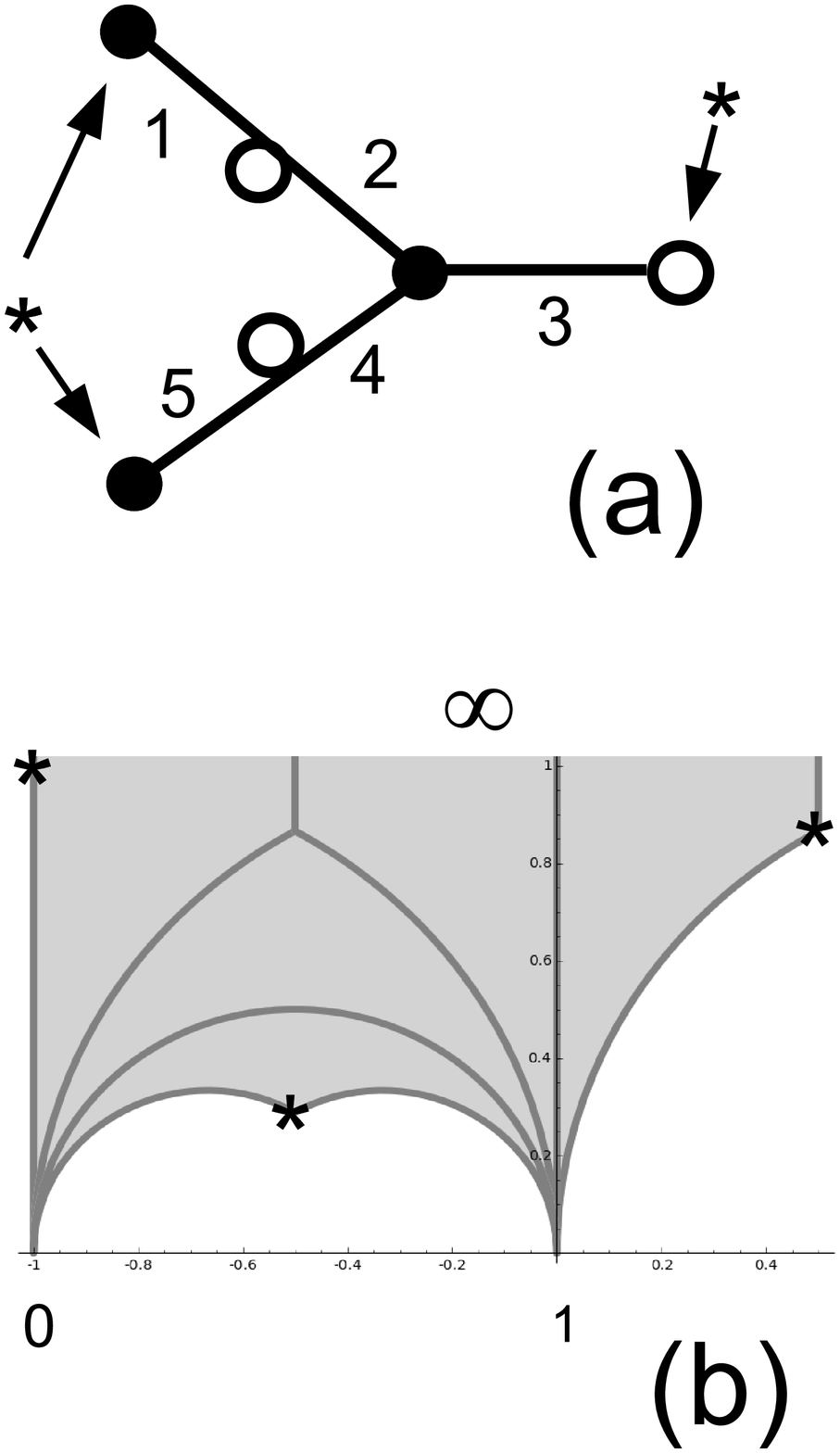}
\includegraphics[width=5cm]{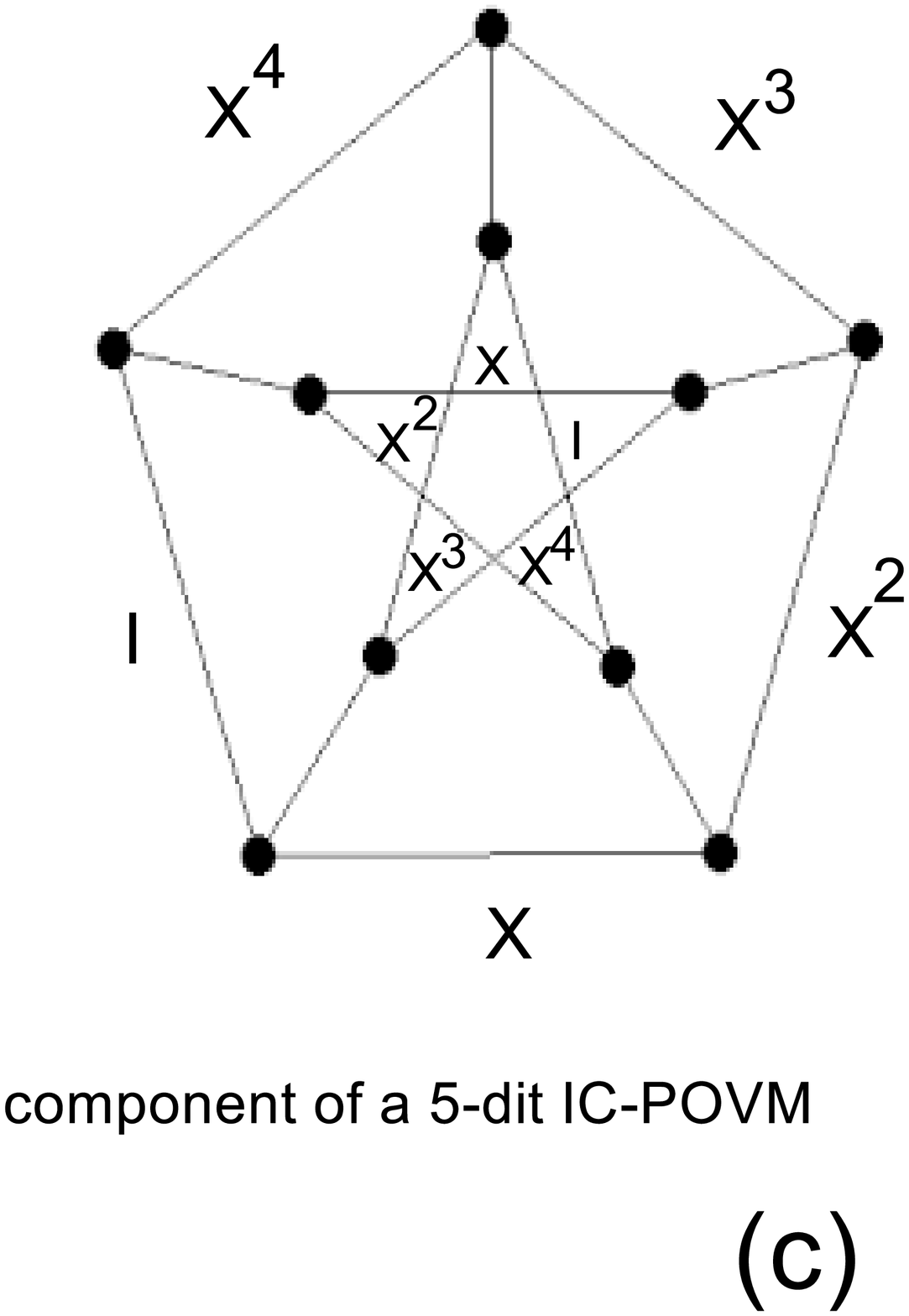}
\caption{Representation of $A_5 \cong  5A^0$ as a dessin d'enfant (a) and as the tiling of the fundamental domain (b). The character * denotes the two elliptic points of order $3$. (c) A one-point intersection graph organizing the lines of the $5$-dit equiangular IC-POVM defined from the triple products  of constant trace $-\frac{1}{4^3}$.
 }
\end{figure}  

There is one single elliptic point of order two, two elliptic points of order three denoted by the symbol *, one cusp at $\infty$ and two fractions at $-1$ and $0$.
The subgroup $\Gamma_s=5A^0$ is generated by three transformations $z \rightarrow -\frac{z+2}{z+1}$, $z \rightarrow -\frac{2z+1}{3z+1}$ and $z \rightarrow \frac{1}{1-z}$. Fixed points of such transformation correspond to elliptic points of order two at $z=-1+i$ and order three at $z=\frac{1+i \sqrt{3}}{2}$ and $z=-\frac{1}{2}+\frac{i}{2 \sqrt{3}}$.

\begin{figure}[h]
\includegraphics[width=5cm]{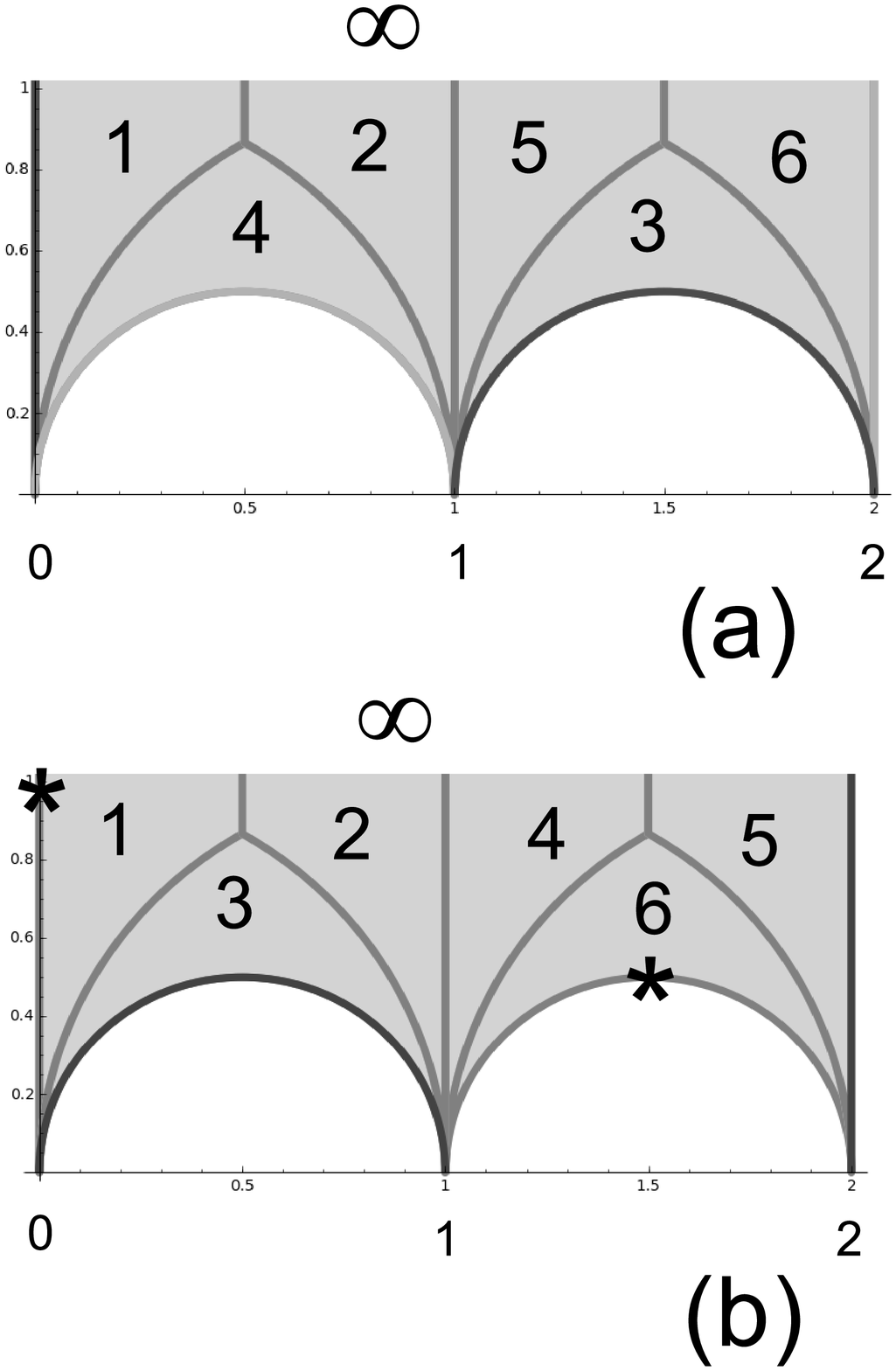}
\includegraphics[width=5cm]{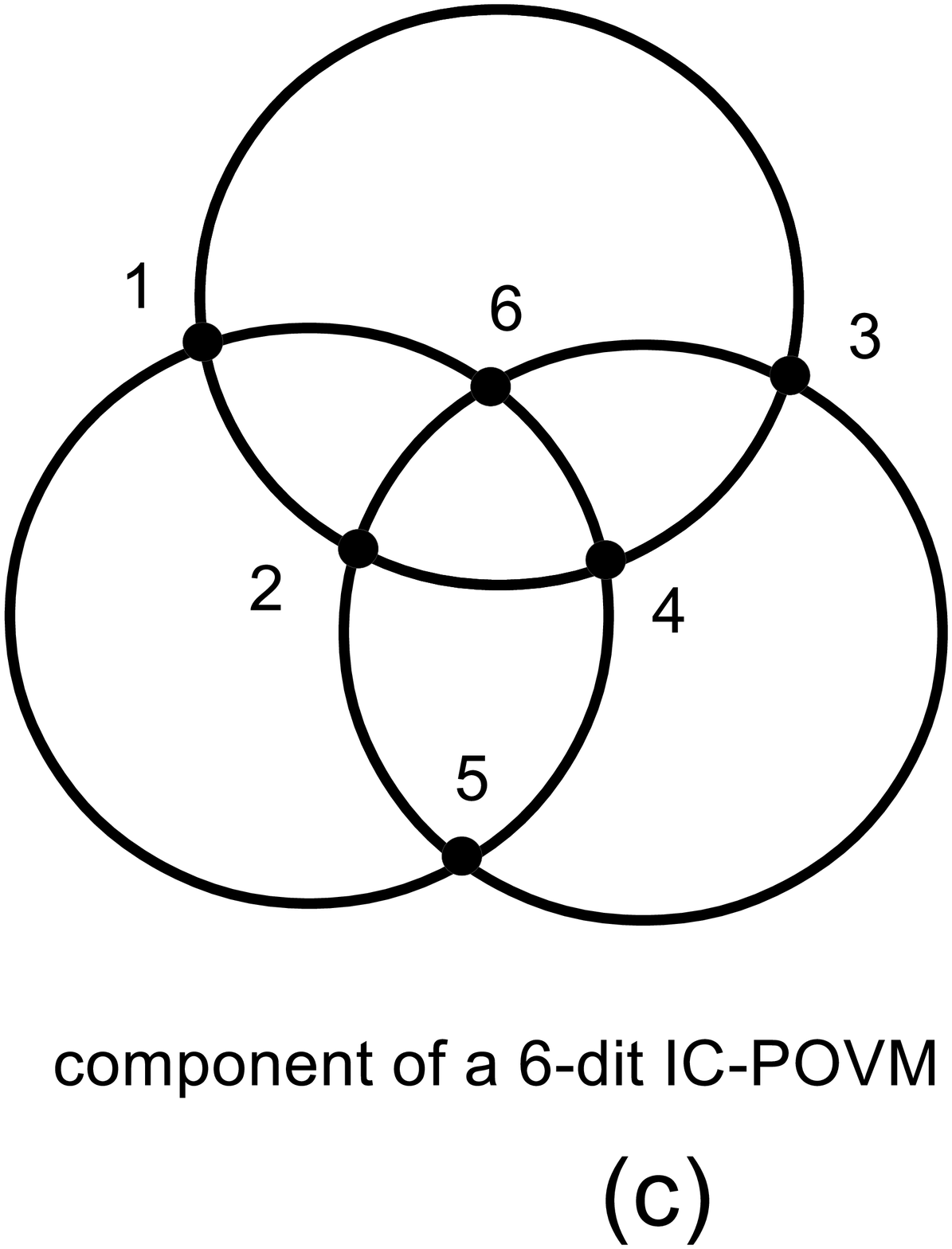}
\caption{(a) Fundamental domain for the genus $1$ group $\Gamma'$, (b) Fundamental domain for the genus $0$ group $6A^1$, the symbol * points out the two elliptic points of order $2$, (c) A basic piece of the $6$-dit IC-POVM with fiducial state of type $(0,1,\omega_6-1,0,-\omega_6,0)$ obtained through the action of Pauli operators $1$ to $6$: the lines correspond  to $4$-tuples products of projectors with constant trace $\frac{1}{9}$ and simultaneously of products equal to $\pm\mathcal{I}$.  There are two disjoint copies looking like Borromean rings with points as \newline $[1\ldots 6]=[\mathcal{I},ZX^3,Z^2,Z^3X^3,Z^4,Z^5X^3]$ (for lines with projector products $-I$) and $[1\ldots 6]=[X^4,Z,Z^2X^3,Z^3,Z^4X^3,Z^5]$ (for lines with projector products $I$).
 }
\end{figure} 

The joined eigenstates of the commuting permutation matrices in $A_5$ that can serve as fiducial states for an IC-POVM are of type $(0,1, 1,1,1)$ and $(0,1,-1,-1,1)$. The latter type allow to construct IC-POVM's such that the pairwise distinct products satisfy  $\left|\left \langle \psi_i \right|\psi_j \right\rangle|^2=\frac{1}{4^2}$, that is the POVM is equiangular with respect to the field norm defined in the introduction. The first type of magic state is dichotomic with values of the products $\frac{1}{4^2}$ and $(\frac{3}{4})^2$. The trace of pairwise products of (distinct) projectors is not constant. For example, with the state $(0,1,-1,-1,1)$, one gets a field norm equiangular IC-POVM in which the trace is trivalued: it is either $1/16$ or $(7\pm 3 \sqrt{5})/32$.

Let us concentrate on the equiangular POVM.
Traces of triple products with constant value $-\frac{1}{4^3}$ define lines organized into a geometric configuration of type $(25_{12},100_3)$. Lines of the configuration have one or two points in common. The two-point intersection graph consists of $10$ disjoint copies of the Petersen graph. One such a Petersen graph is shown in Fig. 3c, the vertices of the graph correspond to the lines and the edges correspond  to the one-point intersection of two lines. As before the labeling is in terms of the operators acting on the magic state.

\subsection{The six-dimensional IC-POVM}

One finds five distinct permutation groups of index $6$ corresponding to subgroups of $\Gamma$ that lead to a six-dimensional IC-POVM. Fiducial states are of type $(0,1,\omega_6-1,0,-\omega_6,0)$ already found in \cite{PlanatGedik} with $\mbox{tr}(\Pi_i \Pi_j)_{i \ne j}=\left|\left \langle \psi_i \right|\psi_j \right\rangle|^2_{i \ne j}=\frac{1}{3}\mbox{or}~~\frac{1}{3^2}$.

The five permutation groups under question are the cyclic group $\mathbb{Z}_6$ leading to congruence subgroups $\Gamma'$ and $\Gamma(2)$, the alternating group $A_4$ leading to the congruence subgroup $3C^0$ \cite{CumminsPauli} and the symmetric group $S_4$ leading to the congruence subgroups $\Gamma_0(4)$ and $\Gamma_0(5)$.
Fundamental domains for $\Gamma'$ and $3C^0$ are shown in Fig. 4a and 4b, respectively.

For a six-dimensional IC-POVM, one discovers a quite simple geometry sustaining the $4$-tuple products of projectors whose product is $\pm I$ and simultaneously have constant trace $\frac{1}{9}$. It consists of two disjoint copies (corresponding to lines whose product of projectors equal $-I$ and $I$, respectively) looking like Borromean rings as shown in Fig. 4).

\subsection{Seven-dimensional IC-POVMs}

Seven-dimensional IC-POVMs with bivalued pairwise products $\left|\left \langle \psi_i \right|\psi_j \right\rangle|^2_{i \ne j}$ are found starting from permutation groups isomorphic to $\mathbb{Z}_7 \rtimes \mathbb{Z}_6$ or $PSL(2,7)$, respectively. The first permutation group corresponds to a non-congruence subgroup. The fiducials of the IC are of type $(0,1,1,1,\pm1 ,\pm 1,\pm 1)$ or $(0,1,-\omega_3-1,\omega_3,1,-\omega_3 -1,\omega_3)$. The second permutation group corresponds to the congruence subgroup $7A^0$ \cite{CumminsPauli}. The fiducials of the IC are of type $(1,0,0,0,1,\pm 1,\pm 1)$ or $(1,0,0,0,i,i,1)$.

\begin{figure}[h]
\includegraphics[width=7cm]{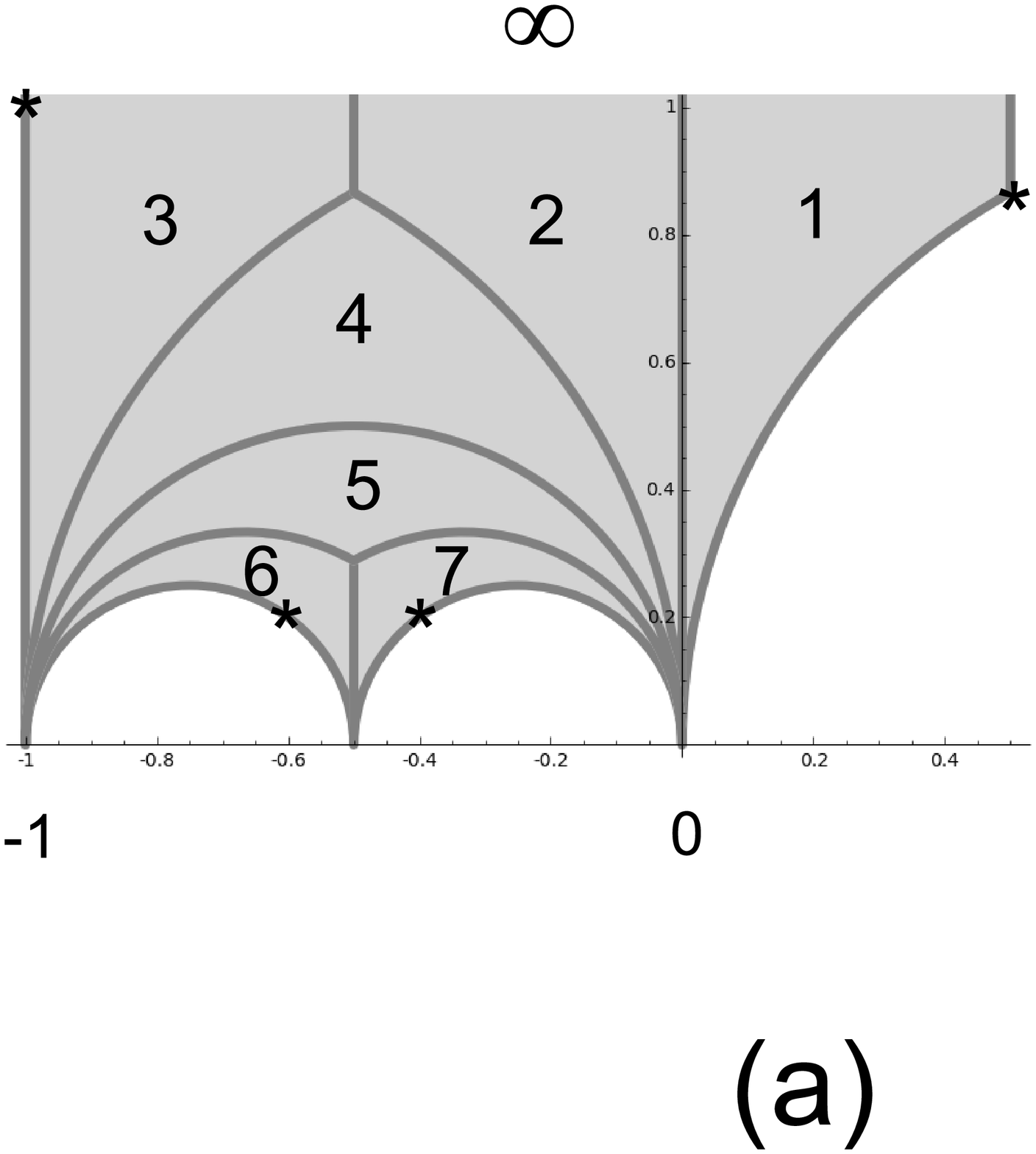}
\includegraphics[width=5cm]{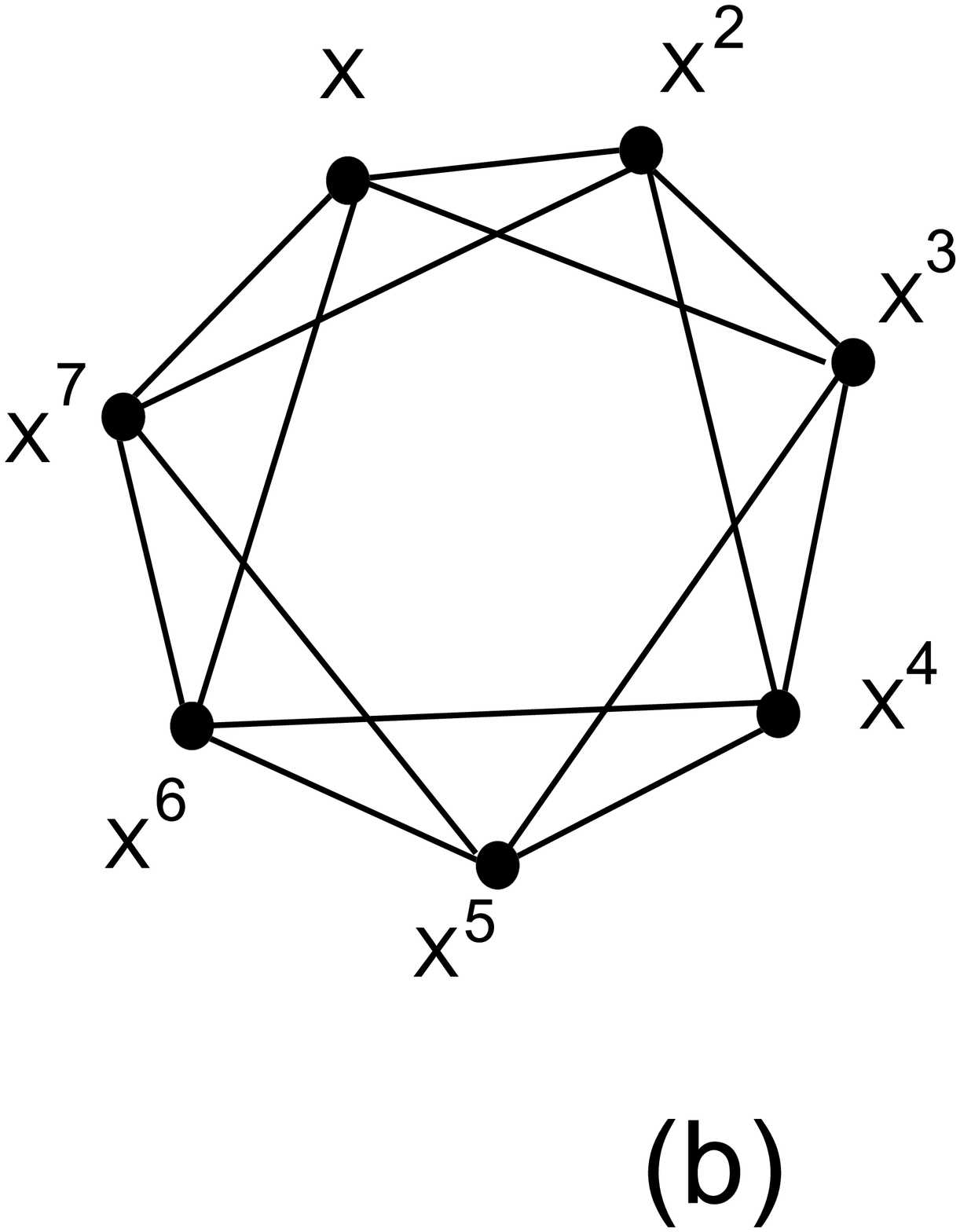}
\caption{ (a) Fundamental domain for the group $7A^0$, (b) A basic component associated to a bivalued $7$-dimensional IC-POVM.
 }
\end{figure} 

The fundamental domain for the group $7A^0$ is shown in Fig. 5a. The geometry of IC triple products is quite complex but one building block may be identified as shown in Fig. 5b (only for ICs with non complex entries in their fiducial).

It may be reminded that an equiangular (with respect to the cyclotomic field norm) $7$-dimensional IC-POVM exists. It is obtained thanks to the group $\mathbb{Z}_7 \rtimes \mathbb{Z}_6$ (in a non-modular representation) using magic permutations \cite{PlanatGedik}. A fiducial such as $(1,-\omega_3-1,-\omega_3,\omega_3,\omega_3+1,-1,0)$ does the job with $\left|\left \langle \psi_i \right|\psi_j \right\rangle|^2_{i \ne j}=\frac{1}{6^{2}}$.

\begin{figure}[h]
\includegraphics[width=6cm]{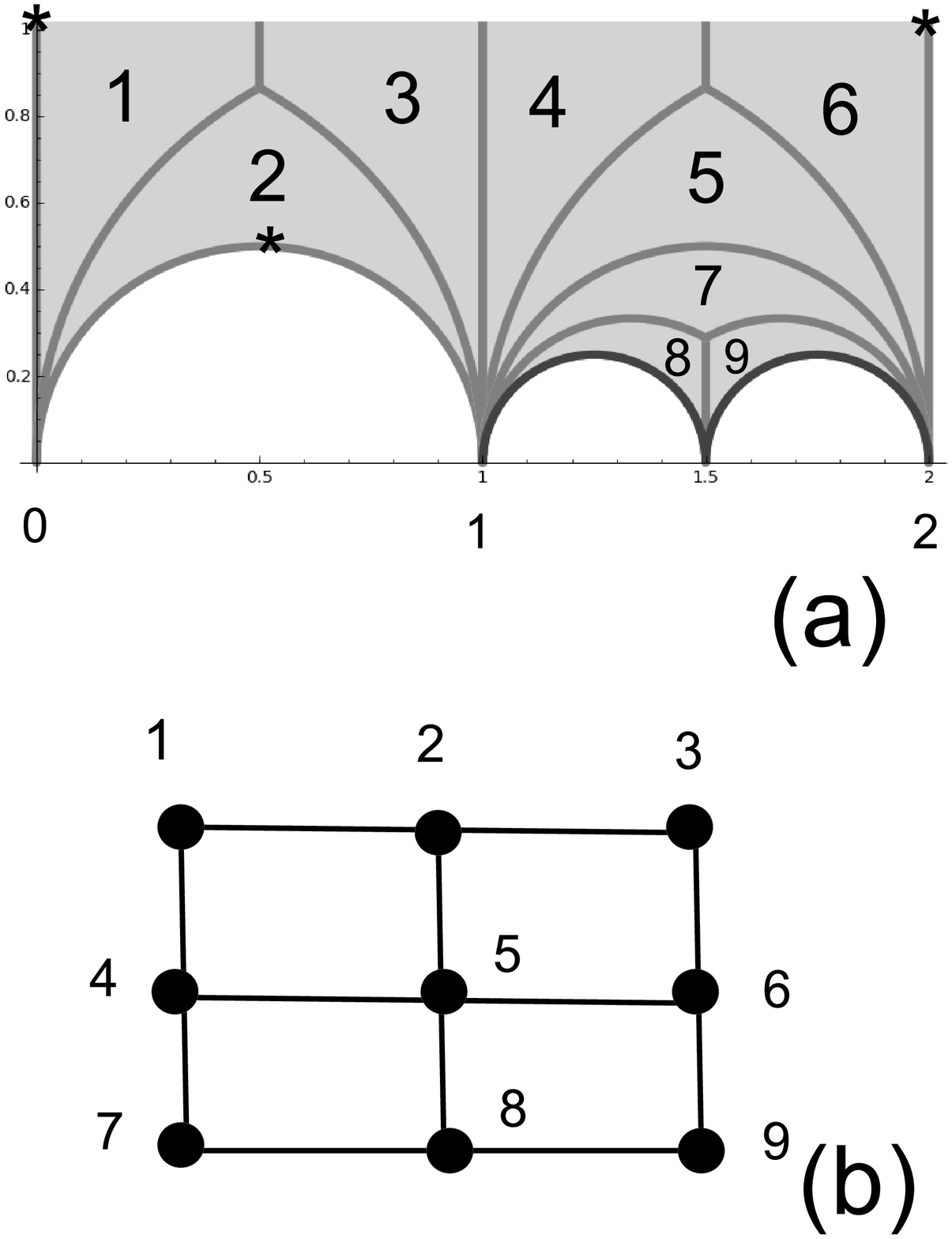}
\includegraphics[width=5cm]{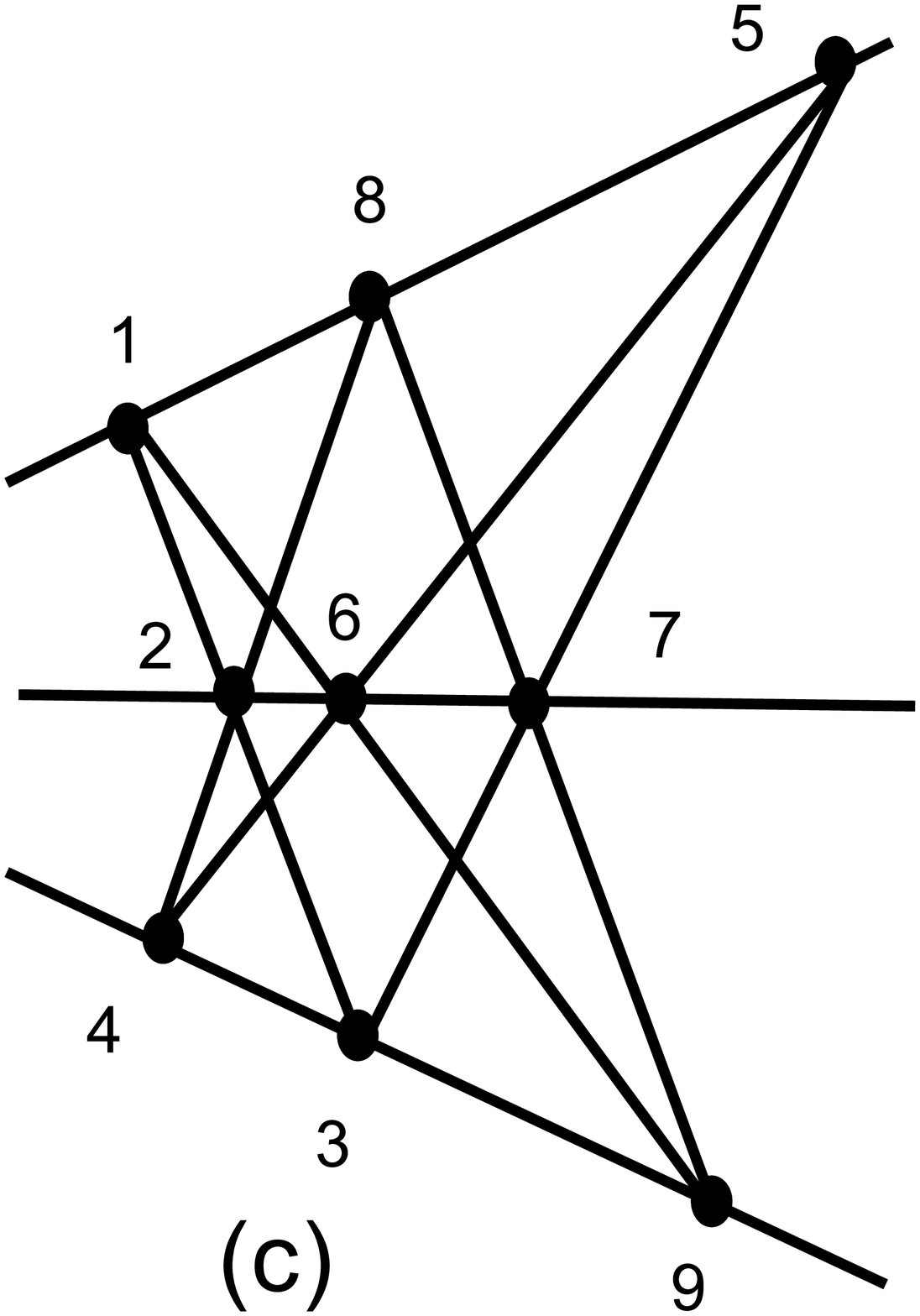}
\caption{ (a) Fundamental domain for the non congruence subgroup of $\Gamma$ associated to the group $\mathbb{Z}_3^2 \rtimes(\mathbb{Z}_2 \rtimes S_4)$, (b) A basic component (a $(3 \times 3)$-grid) in the geometry of the bivalued $9$-dimensional IC-POVM with fiducial $(1,0,0,0,1,0,1,1,0)$. Vertices of the grid are for observables are the two-qutrit observables $[1 \cdots 9]=[I\otimes X,I\otimes XZ, I \otimes XZ^2,Z \otimes X,Z \otimes XZ, Z \otimes XZ^2, Z^2 \otimes X, Z^2 \otimes XZ, Z^2 \otimes XZ^2]$, (c) A basic component (a Pappus configuration) in the geometry of the bivalued $9$-dimensional IC-POVM with fiducial $(1,0,0,0,-1,0,-1,1,0)$. The points are $[1\cdots 9]=[I \otimes Z, I \otimes XZ, I \otimes (XZ^2)^2, Z \otimes I, Z \otimes X,
Z\otimes X^2, Z^2 \otimes Z^2, Z^2 \otimes (XZ)^2, Z^2 \otimes XZ^2]$.
 }
\end{figure} 

\subsection{The three-qubit Hoggar SIC}
The approach based on permutation groups fails to identify any IC-POVM, see \cite{Stacey2016} for details about the geometry of the Hoggar SIC in relation to its covariance under the three-qubit Pauli group. A noticeable result is that the three-qubit SIC embeds the dual of the generalized hexagon $GH(2,2)$ \cite[Fig. 3]{PlanatGedik}. The latter geometry is connected to the eight-dimensional Kochen-Specker theorem.

\subsection{Nine-dimensional IC-POVMs}
The smallest permutation group isomorphic to a subgroup of $\Gamma$ and useful to build a nine-dimensional IC-POVM is $P=\left\langle (1,2,3)(4,5,6)(7,8,9),(3,4)(5,7)(8,9) \right\rangle$$ \cong \mathbb{Z}_3^2 \rtimes(\mathbb{Z}_2 \rtimes S_4)$. The corresponding fundamental domain is shown in Fig. 6a, the subgroup is non-congruence and contains three elliptic points of order two at $i$, $2+i$ and  $(1+i)/2$ and two cusps at $\frac{3}{2}$ and $\infty$. The related IC-POVM has bi- or three-valued distinct pairwise products. Let us choose the fiducial state for the bivalued case as $(1,0,0,0,\pm 1,0,0,\pm 1,1,0)$. For the state with positive entries and traces of triple products equal to $-\frac{1}{8}$, the geometry consists of $6$ copies of a $3 \times 3$ grid as shown on Fig. 6b. The products of the observables on a row or a column of the grid equal $1$, $\omega_3$ or $\omega_3^2$. For the state with positive and negative entries, and traces of triple products equal to $\frac{1}{8}$, the geometry consists of $9$ copies of the Pappus configuration as shown in Fig. 6c. It allows a proof of the $2$QT Kochen-Specker theorem \cite[Sec. 2.7 and Fig. 5]{PlanatGedik}.

A trivalued IC-POVM follows from the permutation group
of order $504$ that is also non congruence. The fiducial is of type $(0,1,1,1,1,1,1,1,1)$.

It is useful to remind that a group generated by two magic permutations and isomorphic to $\mathbb{Z}_3^2 \rtimes \mathbb{Z}_4$ has been used to construct a bivalued IC-POVM starting from a fiducial of type $(1,1,0,0,0,0,-1,0,-1)$ \cite{PlanatGedik}. Noticeably, It contains the Pappus configuration in the organization of its triple products.

\small
\begin{table}[h]
\begin{center}
\begin{tabular}{|l|c|r|l|}
\hline 
\hline
dim & subgroups of $\Gamma$ leading to an IC-POVM  & pp & geometry\\
\hline
\hline
2  & none &  1 & tetrahedron \cite{Bravyi2004} \\
\hline
3  & $\Gamma_0(2)$ & $1$ & Hesse SIC \cite{Bengtsson2010} \\
\hline
4  & under 2QB Pauli group &   \\
  & $\Gamma_0(3), 4A^0$ & 2 &  $GQ(2,2)$ \\
\hline
5  & $5A^0$ & 1 & Petersen graph \\
\hline
6  & $\Gamma', \Gamma(2), 3C^0,\Gamma_0(4),\Gamma_0(5)$ & 2 & Borromean ring \\
\hline
7  &  $7A^0$ & 2 & Fig. 5b \\
  & NC$(0,6,1,1,[1^1 6^1])$ & 2 &  \\
  & none & 1 & \cite{PlanatGedik} \\
\hline
8  & none under 3QB, 8-dit, 4-dit-QB Pauli group & 1 & Hoggar SIC \cite{Stacey2016,PlanatGedik} \\
\hline
9  & under 2QT Pauli group &   \\
  & NC$(0,8,3,0,[1^1 8^1])$ & 2 & $(3 \times 3)$-grid, Pappus \\
  & NC$(0,9,1,3,[9^1])$ &3 & $[81_8,216_3]$  \\
\hline
\hline
10  & $5C^0$ & 5& \\
\hline
11  & $11A^0$ & 3 & $[11_3]$
 \\
\hline
 12 & under 2QB-QT Pauli group  &  &  \\
  & $10A^1$ & 5 & $K(3,3,3,3)$ \\
  & NC $(0,8,4,0,[4^1 8^1])$  & 5 & Hesse ($\times 16$) \\
  & NC$(0,8,4,0,[4^1 8^1])$  & 6 & $[48_7,112_3]$
	\\
	\hline
	 12 & under 12-dit Pauli group  &  &  \\
	 & $8A^1$,  NC$(0,8,4,0,[4^18^1])$ & 11,7 &  \\
\hline	
13  &NC$(0,6,1,1,[ 1^1 6^2])$ & 4 & \\
\hline
14  &$7C^0$, NC$(0,6,0,2,[ 1^1 6^2])$, $14A^1$ & 12,5,6 & \\
\hline
15  & $5E^0$, NC$(0,6,3,0,[ 3^1 6^2])$, $15A^1$, $10B^1$  & 5,4,10,3 & \\
\hline
16  & none under 4QB and 2 4-dit Pauli group & &  \\
\hline
 18 & under 18-dit or 2QT-QB Pauli group  &  &  \\
 & $\Gamma_0(10)$, NC$(1,8,0,0,[ 2^1 8^2])$          & 7,5 & \\
\hline
 19  &NC$(0,6,1,1,[ 1^1 6^3])$ & 3 & \\
\hline
21  &NC$(0,6,3,0,[ 3^1 6^3])$, NC$(0,6,1,0,[ 1^1 2^1 6^3])$  & 4 & \\
  &NC$(0,14,7,0,[ 7^1 14^1])$, NC$(0,8,3,0,[ 1^1 4^1 8^2])$  & 59,4 & \\
	\hline
24  & none under 3QB-QT Pauli group  &  & \\
\hline
24  & under 24-dit Pauli group &  & \\
  & $24A^1$, NC$(2,12,0,0,[12^2])$, $20B^1$, $12F^1$  &40,56,40,30  &  \\
  & NC$(0,6,2,0,[3^26^3])$, NC$(1,8,0,0,[4^28^2])$ &8,7  & \\
	&  $21A^2$, $24A^1$  &23,60  & \\
\hline
25  &  under 25-dit Pauli group  &  & \\
  &NC$(0,10,5,1,[5^1 10^2])$ & 15 & \\
\hline
27  &  under 3QT Pauli group  &  & \\
  &NC$(0,6,1,0,[1^1 2^1 6^4])$, NC$(0,8,3,0,[1^1 2^1 8^3])$ & 4 & Pappus\\
\hline
\hline
\end{tabular}
\caption{A summary of the subgroups of the modular group $\Gamma$ (column 2) allowing the construction of IC-POVMs in the corresponding dimension (column 1). When non-congruence the signature NC$(g,N,\nu_2,\nu_3,[c_i^{W_i}])$ is made explicit. Column $3$ shows the minimal number $pp$ of pairwise distinct products needed. Column 4 features the  underlying geometries when recognized.}
\end{center}
\end{table}
\normalsize 

\subsection{Higher dimensional IC-POVMs}

IC-POVMs found in dimension $d$ higher than $9$ are summarized in the second half of Table 2. The minimal number of pairwise products needed increases with $d$ (as in column 3) and the subgroup $\Gamma_s$ occurring in the construction is quite often non congruence (as expressed in column 2). The symmetry underlying triple products of projectors is not simple and not easily recognizable. We could not find (modular group based) IC-POVMs in dimensions $d=8,16,17,22,23$. In dimension $24$, the found ICs are covariant under the $24$-dit Pauli group, not under the $3QB$-$QT$ group. Finally, in dimension $27$, one finds a $4$-valued IC-POVM, covariant under the $3$QT Pauli group, whose structure of the triple products consists of 81 copies of the Pappus configuration. 

Table 1 summarizes the results obtained so far.

\section{Conclusion}
{\it It would be nice if we could design a virtual reality in Hyperbolic Space, and meet each other there} \cite{Knuth}.

The continuing search of mathematical structures governing the weirdness of quantum theory is catalyzed by applications. A universal quantum computer needs non-stabilizer states, i.e. states that are not eigenstates of a Pauli group. The finding of distillable qubit magic states in \cite{Bravyi2004} prompted us to pass to higher dimensions with the methods of permutation theory \cite{PlanatRukhsan}. It was soon observed that an interesting subset of magic states could be seen as fiducials for minimal IC-POVMs of the corresponding dimension \cite{PlanatGedik}. Now, the present work connects universal quantum computing, IC-POVMs, quantum contextuality and the subgroups of finite index $\Gamma_s$ of the modular group $\Gamma$. This allows to see the magic/fiducial states arising from appropriate permutation gates in the new language of fundamental domains and their copies under the discontinuous action of $\Gamma_s$. Thus and unexpectedly, a short circuit occurs between $\Gamma$ and quantum theory that has not be investigated so far. The group $\Gamma$ is the starting point of the modularity theorem that connects  elliptic curves over the rationals and modular forms. A jewel of mathematics is tethered to our best physical theory.

\section*{Acknowledgements}
The first author acknowledges his student Etienne Donier-Meroz for help in identifying the elliptic points and cusps of subgroups of the modular group in the Sage software. He also thanks Alain Giorgetti for his check of Magma calculations.

\section*{Funding}

This work was supported by the French \lq\lq Investissements d'Avenir" program, project ISITE-BFC (contract ANR-15-IDEX-03).

\end{document}